\documentclass[prd,preprint,showpacs,preprintnumbers,amsmath,amssymb,amsthm,floatfix]{revtex4}
\usepackage{amsthm}
\usepackage{graphicx}
\usepackage{natbib}
\usepackage{bm}
\usepackage{mathrsfs}
\usepackage{verbatim}
\usepackage{wasysym}

\newcommand{\hf}{{\frac{1}{2}}}
\def\d{\mathrm{d}}

\begin{document}

\title{On Event Horizons in De Sitter Yang-Mills Theory}
\author{Timothy D. Andersen}
\email{andert@alum.rpi.edu}
\date{Received: date / Accepted: date}
\begin{abstract} De Sitter Quantum Gravity is a Yang-Mills theory based on the de Sitter or SO(4,1) group and a promising candidate for a quantum theory of gravity. In this paper, an exact, static, spherically symmetric solution of the classical equations is derived. I show that when the Schwarzchild radius to distance ratio is at post-Newtonian order the theory agrees with general relativity for all parameters but that, once the ratio becomes closer to unity, they differ. At the Schwarzchild radius from a black hole singularity, general relativity predicts an event horizon, which has become a controversial topic in quantum gravity because of information preservation issues. In the De Sitter theory I show, however, that time-like escape paths exist for any mass black hole until the singularity itself is reached. Since an event horizon has never been directly observed and there is currently no observation on which the two theories disagree, this provides a powerful test of the De Sitter theory.
\end{abstract}

\maketitle
De Sitter Quantum Gravity was proposed as a spin-1, Yang-Mills theory of quantum gravity on a Minkowski metric, $(-+++)$, that, while agreeing with all observations of gravitational phenomena to date, could also be perturbatively renormalized. Black holes have not been studied in the context of the theory, nor have any exact solutions to the equations been found other than for the accelerating expansion of a Robertson-Walker universe \cite{Andersen:2013}. This issue is particularly relevant because (1) information preservation in black holes is a source of present controversy and (2) black holes are being discovered more frequently.

The literature on black holes is extensive going back to the 18th century with the notion of a ``dark star'' in Newtonian gravity\cite{Michell:1784}\cite{Israel:1989} to Schwarzchild's exact solution to the Einstein Field Equations in 1916 \cite{Schwarzchild:1916}. The Schwarzchild radius was identified as the radius of a one-directional ``membrane'' around a non-rotating, neutral black hole in 1958 \cite{Finkelstein:1958}, and, although purely speculative since that time it has been a major source of research in the well-accepted framework of general relativity (GR). Controversy over black holes erupted, however, when Hawking suggested that black holes evaporate under a quantum process of releasing Hawking radiation from just outside their event horizons \cite{Hawking:1975} and further predicted that quantum information could be lost within the black hole as a result---a loss of unitarity that threatened the basis of quantum mechanics \cite{Preskill:1992}. Although many physicists, including Hawking have come to accept that information can be released from black holes, e.g., mangled as Hawking radiation, the fact remains that event horizons are speculative objects, not yet observed, and nothing can be confirmed about any of these hypotheses. Although black holes have been inferred to exist from X-ray sources \cite{Celotti:1999} and at the center of galaxies\cite{Kormendy:1995}\cite{King:2003}, direct observation of the event horizon has not been possible. Having observations of these very strong gravitational phenomena would provide some of the best possible evidence for and against alternatives to general relativity that deviate from it classically but only in gravitational fields stronger than have been observed. De Sitter quantum gravity is one such theory.

The starting point for a study of black holes is the static, spherically symmetric solution to the classical equations. In this paper, an exact such solution to the classical Yang-Mills equations is found. This solution is shown to be equivalent to that of general relativity to post-Newtonian order which is sufficient for all precise observations to date. Close to the Schwarzchild radius from a black hole, however, De Sitter theory and general relativity deviate significantly. While general relativity predicts an event horizon at which the escape velocity equals the speed of light, the De Sitter theory predicts that escape velocity only reaches the speed of light at the singularity, indicating that black holes, while (classically) emitting no light, have zero-radius event horizons. This prediction is important because currently there is no test that can separate the predictions of De Sitter Yang-Mills theory from those of general relativity. (These include all the classical observations including N-body motion, redshift, time dilation, periastron precession, gravitational radiation, and the strong equivalence principle, which are derived in \cite{Andersen:2013}.) The observation or non-observation of the predicted event horizon would falsify one or the other.

\section{De Sitter Yang-Mills Theory Definition}

In De Sitter Yang-Mills theory, observer frames are gauges of the Yang-Mills theory, with the matrix potential $A_\mu = G_{\mu\nu}V^{\nu} + \hf H_{\mu\nu\lambda}M^{\mu\nu}$ where $V^\nu$ and $M^{\mu\nu}$ are generators of the de Sitter Lie algebra. Curvature in spacetime is represented by the $G_{\mu\nu}$ potential while torsion (twisting) is represented by the $H_{\mu\nu\lambda}$ potential. The action is the standard Yang-Mills action for the de Sitter group,
\begin{equation}
S_{gravity} = -\frac{1}{4g^2}\int \d^4x\, E_{\mu\nu\lambda} E^{\mu\nu \lambda} + \hf F_{\mu\nu\alpha\beta} F^{\mu\nu\alpha\beta}.
\label{eqn:action}
\end{equation}  Here the coupling constant is the non-dimensional $a_g=g^2$,
\begin{eqnarray}
 E_{\mu\nu\lambda} = \partial_\mu G_{\nu\lambda} - \partial_\nu G_{\mu\lambda} +\nonumber\\ \eta^{\sigma\rho}\left(G_{\mu\lambda}H_{\nu\sigma\rho} - G_{\mu\rho}H_{\nu\sigma\lambda} - G_{\nu\lambda}H_{\mu\sigma\rho} + G_{\nu\rho}H_{\mu\sigma\lambda}\right),
\label{eqn:transforce}
\end{eqnarray}
and
\begin{eqnarray}
F_{\mu\nu\alpha\beta} = \partial_\mu H_{\nu\alpha\beta} - \partial_\nu H_{\mu\alpha\beta} + G_{\mu\alpha}G_{\nu\beta} - G_{\mu\beta}G_{\nu\alpha} + \Phi_{\mu\nu\alpha\beta} - \Phi_{\mu\nu\beta\alpha}
\label{eqn:rotforce}
\end{eqnarray}  where
\begin{eqnarray}
\Phi_{\mu\nu\alpha\beta}  = \hf\eta^{\sigma\rho}\left(H_{\mu\sigma\alpha}H_{\nu\rho\beta} - H_{\mu\sigma\alpha}H_{\nu\beta\rho} -\right.\nonumber\\ \left.H_{\mu\alpha\sigma}H_{\nu\rho\beta} + H_{\mu\alpha\sigma}H_{\nu\beta\rho}\right).
\end{eqnarray} Indexed are raised and lowered with the Minkowski metric.

The quantum field theory is given by,
\[
Z = e^{iS_{gravity}/\hbar}.
\] 

\section{Classical Equations}
The classical equations of the Yang-Mills theory are:
\begin{eqnarray}
\label{eqn:motion1}
 \partial^\mu E_{\mu\nu\lambda} + \eta^{\sigma\rho}\left(E_{\mu\nu\lambda}H^\mu{}_{\sigma\rho} - E_{\mu\nu\rho}H^\mu{}_{\sigma\lambda} -\right.\nonumber\\\left. G^\mu{}_{\lambda}F_{\mu\nu\sigma\rho} + G^\mu{}_{\rho}F_{\mu\nu\sigma\lambda}\right) & = & -8\pi a_gJ_{\nu\lambda}\\
\label{eqn:motion2}
\partial^\mu F_{\mu\nu\alpha\beta} + E_{\mu\nu\alpha}G^\mu{}_{\beta} - E_{\mu\nu\beta}G^\mu{}_{\alpha} +\Sigma_{\nu\alpha\beta} - \Sigma_{\nu\beta\alpha} & = & -8\pi a_gS_{\nu\alpha\beta}
\end{eqnarray}  where
\begin{eqnarray}
\Sigma_{\nu\alpha\beta} = \hf\eta^{\sigma\rho}\left(F_{\mu\nu\sigma\alpha}H^\mu{}_{\rho\beta} - F_{\mu\nu\sigma\beta}H^\mu{}_{\beta\rho} - \right.\nonumber\\\left.F_{\mu\nu\alpha\sigma}H^\mu{}_{\rho\beta} + F_{\mu\nu\alpha\sigma}H^\mu{}_{\beta\rho}\right).
\end{eqnarray}

Boundary conditions are: 
$
G_{\mu\nu}\rightarrow\eta_{\mu\nu},\quad \partial_\lambda G_{\mu\nu} \rightarrow 0
$ and 
$
H_{\mu\nu\lambda}\rightarrow 0,\quad \partial_\alpha H_{\mu\nu\lambda} \rightarrow 0
$ as $x_\mu\rightarrow\infty$.

Gauge transformations can be performed in component form. Consider the $SO(4,1)$ de Sitter transformation $\Delta(x) = \xi_\mu(x) V^\mu + \hf \chi_{\mu\nu} M^{\mu\nu}$. The gauge transformation is 
\begin{equation}
G'_{\mu\nu} = G_{\mu\nu} + \partial_\mu \xi_\nu(x) + \eta^{\rho\lambda}(\xi_\lambda(x)H_{\mu\rho\nu} - \xi_\lambda(x)H_{\mu\nu\rho} + \chi_{\lambda\nu}(x)G_{\mu\rho} - \chi_{\nu\rho}G_{\mu\lambda})
\label{eqn:Ggauge}
\end{equation} and
\begin{eqnarray}
H'_{\mu\nu\lambda} & = & H_{\mu\nu\lambda} + \partial_\mu \chi_{\nu\lambda}(x) + \xi_\nu(x)G_{\mu\lambda} - \xi_\lambda(x)G_{\mu\nu} + \nonumber\\ & &\hf\eta^{\kappa\rho}(\chi_{\kappa\nu}(x)H_{\mu\rho\lambda} - \chi_{\kappa\nu}(x)H_{\mu\lambda\rho} - \chi_{\nu\kappa}(x)H_{\mu\rho\lambda} + \chi_{\nu\kappa}(x)H_{\mu\lambda\rho}),
\end{eqnarray} which can be found from the Lie algebra.

For discrete test particles of mass $m$, the action,
\begin{equation}
\mathcal{S} = -\frac{1}{2m}\int d\sigma G_{\mu\nu} J^{\mu \nu},
\label{eqn:lagr2}
\end{equation} where $J^{\mu\nu}$ is the test particle current with path $x^\mu(\sigma)$, is minimized to produce the standard geodesic equation:
\begin{equation}
\frac{d^2x^\lambda}{d\tau^2} + \frac{1}{2}(G^{-1})^{\lambda\nu}\left[\partial_\rho G_{\mu\nu} + \partial_\mu G_{\rho\nu} -  \partial_\nu G_{\mu\rho}\right]\frac{dx^\mu}{d\tau}\frac{dx^\rho}{d\tau} = 0,
\label{eqn:geo}
\end{equation} where $\tau$ is proper time. Note: there is a distinction between raising indexes and inverting $G_{\mu\nu}$. Its inverse is $(G^{-1})^{\mu\nu}$ (such that $(G^{-1})^{\mu\alpha}G_{\alpha\nu} = \delta^\mu_\nu$) not $G^{\mu\nu} = \eta^{\mu\alpha}\eta^{\nu\beta}G_{\alpha\beta}$.

The classical coupling of the De Sitter gravitational field to the Dirac action is relatively simple. Let
\begin{equation}
S_{Dirac} = \int d^4x\,i\bar{\psi}\gamma^\mu(\partial_\mu + im_gA_\mu)\psi
\end{equation} be the Dirac action and $A_\mu$ a matrix potential for the De Sitter group with $m_g$ the mass quantum. Break the matrix potential into a sum over de Sitter generators for spinor fields,
\begin{equation}
iA_\mu = G_{\mu\nu}P^\nu + \frac{1}{2}H_{\mu\nu\lambda}\sigma^{\nu\lambda}.
\end{equation} Now apply the Euler-Lagrange equation to the Lagrangian in $S_{Dirac}$ with respect to $G_{\mu\nu}$ and $H_{\mu\nu\lambda}$ and we get expressions for energy and spin-density, $J_{\mu\nu}$ and $S_{\mu\nu\lambda}$. Because $\gamma^\mu\gamma^\mu = \eta^{\mu\mu}$ and, otherwise, $\gamma^\mu\gamma^\nu=-\gamma^{\mu}\gamma^\nu$ when $\mu\neq\nu$, the energy density simplifies to \cite{Zee:2003},
\begin{equation}
J^{\mu\nu} = -\frac{m_g}{2}\bar{\psi}(\delta^{\mu\nu} - iP^{\mu\nu})\psi,
\end{equation} where $-P^{i0} = P^{0i} = -\sigma^{0i}$ and $P^{ij} = \sigma^{ij}$. The off-diagonals of $J^{\mu\nu}$ represent the contribution of total angular momentum to the energy density. If the off-diagonals are zero, with no angular momentum, then
\begin{equation}
J_{\mu\nu} = -\frac{m_g}{2}\bar{\psi}\psi \delta_{\mu\nu}
\end{equation} This simplifies, in classical equations, to
\begin{equation}
J_{\mu\nu} = -m\delta_{\mu\nu}
\end{equation}

\section{Spherically Symmetric Vacuum Solution}
Now the spherically symmetric vacuum solution will be found. The solution gives the result in isometric coordinates. Let coordinates be spherical $(t,r,\theta,\phi)$ with $r$ radius, $\theta$ colatitude, $\phi$ azimuth. Consider a static, spherically symmetric space such that $\partial G_{\mu\nu}/\partial t = 0$ and $\partial H_{\mu\nu\lambda}/\partial t = 0$. The following coordinate transformations $t\rightarrow -t$, $\phi\rightarrow -\phi$, and $\theta\rightarrow -\theta$ (but not $r\rightarrow -r$ since $r\geq 0$) are required to be invariant. The allowable non-zero components of $G$ are 
\begin{equation}
G_{tt} = A(r),\quad G_{rr} = B(r),\quad G_{\theta\theta} = B(r)r^2,\quad G_{\phi\phi} = B(r)r^2\sin^2\theta.
\label{eqn:G}
\end{equation} For $H$, 
\begin{equation}
H_{ttr} = C(r) = -H_{trt},\quad H_{\theta\theta r} = D(r)r^2 = -H_{\theta r\theta},\quad H_{\phi\phi r} = D(r)r^2\sin^2\theta = H_{\phi r \phi}.
\label{eqn:H}
\end{equation} 

For any function $f$, let $\dot{f} = \partial f/\partial r$. The classical equations \ref{eqn:motion1} become:
\begin{eqnarray}
\ddot{A} + \dot{B}C + B\dot{C} + 2A/r + 2BC/r & = & 0,\nonumber\\
\dot{B}/r + BD/r & = & 0,\nonumber\\
\ddot{B}r^2 + r\dot{B} + \dot{B}D r^2 + B\dot{D}r^2 + rBD & = & 0,\nonumber
\end{eqnarray} and \ref{eqn:motion2} gives
\begin{eqnarray}
2\ddot{C} - 4C/r^2 + 4\dot{C}/r + B\dot{A} + B^2C & = & 0,\nonumber\\
2\ddot{D}r^2 + 4r\dot{D} - 4D + r^2B\dot{B} + r^2B^2D & = & 0,\nonumber
\end{eqnarray}

These have the following solutions 
\begin{eqnarray}
  D \left( r \right) & = &{\frac {{ C_2}}{{r}^{2}}},\nonumber\\
 C \left( r \right) & = &{ C_4}\,r+{\frac {{ C_5}}{{r}^{2}}}+{ C_6}\, \left( {r}^{3}\int \!{e^{{\frac {{ C_2}}{r}}}}{r}^{-2}{dr}-\int \!r{e^{{\frac {{ C_2}}{r}}}}{dr} \right) {r}^{-2},\nonumber\\
 A \left( r \right) & = &\int \!- \Bigg( C_3 {r}^{3}{ C_6}\,\int \!{e^{{\frac {{ C_2}}{r}}}}{r}^{-2}{dr}\,{e^{{\frac {{ C_2}}{r}}}}-\nonumber\\ & &C_3 { C_6}\,\int \!r{e^{{\frac {{ C_2}}{r}}}}{dr}\,{e^{{\frac {{ C_2}}{r}}}}+6\,{ C_6}+{r}^{3} C_3{ C_4}\,{e^{{\frac {{ C_2}}{r}}}}+{ C_3 C_5}\,{e^{{\frac {{ C_2}}{r}}}} \Bigg) {r}^{-2}{dr}+{ C_7},\nonumber\\
 B \left( r \right) & = & C_3{e^{{\frac {{ C_2}}{r}}}}  
\end{eqnarray} where one parameter, $C_1=0$, has been removed for readability. To match general relativity to the necessary order, let $C_2=-r_s=C_5$, $C_3=-1$, $C_7=0$, $C_4,C_6=0$ the final solution is
\begin{equation}
A(r) = -e^{-r_s/r},\quad B(r) = -e^{-r_s/r},\quad C(r) = D(r) = -r_s/r^2
\end{equation} where $r_s$ is the Schwarzchild radius $r_s=2Gm/c^2$ for a body of mass $m$. For $r_s/r\ll 1$, $A(r)\approx -1 + r_s/r - (r_s/r)^2 + O[(r_s/r)^3]$ and $B(r) \approx -1 + r_s/r + O[(r_s/r)^2]$. Boundary conditions require $\lim_{r\rightarrow\infty} A(r) = -1$, $\lim_{r\rightarrow\infty} B(r) = 1$ and $\lim_{r\rightarrow\infty} C(r) = 0$. Thus, a simple gauge transformation applying \ref{eqn:Ggauge} with $\xi_r = 2r$ adds $2$ to $B(r)$ so that $B(r)\approx 1 + r_s/r + O[(r_s/r)^2]$. Plugging these into the potential \ref{eqn:G} approximates the isotropic spherically symmetric static potential of general relativity to post-Newtonian order with post-Newtonian parameters $\beta=\gamma=1$ and all others zero. The general case for N-bodies is proved in \cite{Andersen:2013}.

In the case of the Einstein Field Equations the Schwarzchild metric is the only spherically symmetric solution up to a change in coordinates. That claim is not proved for these equations but is conjectured to be true.

\section{Results}
The post-Newtonian order is the highest order to which current observations of non-cosmological phenomena have been directly observed, e.g., one of the strongest being the double binary pulsar system PSR J0737-3039A/B \cite{Kramer:2006}. Currently, black holes have only been observed indirectly as massive dark objects, particularly at the center of galaxies \cite{Kormendy:1995}\cite{King:2003}. The observation of black holes more directly than current technology permits promises to usher in a new age of strong gravity research. It will also provide a test capable of falsifying either general relativity or the De Sitter theory. This hinges on the existence of the event horizon, a never observed prediction of general relativity that shields a black hole's singularity and allows nothing that crosses it to escape \cite{Finkelstein:1958}. 

The event horizon for a black hole in general relativity is at the radius $r=r_s/4$. Let units be such that $c=1$. The isotropic metric, $g_{\mu\nu}$, is given by the following \cite{Misner:1973}:
\[
d\tau^2 = -\frac{(1-\frac{r_s}{4r})^2}{(1+\frac{r_s}{4r})^2}dt^2 + (1+\frac{r_s}{4r})^4(dr^2 + r^2(d\theta^2 + \sin^2\theta d\phi^2)),
\] valid for $r\ge r_s/4$. When $r\sim r_s$, the potential $G_{\mu\nu}$ diverges significantly from that of general relativity. While time dilation and redshift become essentially infinite at the event horizon in GR where $g_{00}=0$ (in either isotropic or Schwarzchild coordinates), in the De Sitter theory nothing special happens at this distance. Rather, time dilation and redshift become infinite at the singularity itself, $r\rightarrow 0$, where $G_{00}\rightarrow 0$. This implies that a light source falling into a black hole would continue to emit redshifted light until it collides with the singularity itself and only then go dark. 

Now we will calculate the escape velocity as a function of the isometric radius, $r$. Consider a stationary test body of mass $m$ at radius $r$ (in isotropic coordinates) above a non-rotating, neutral black hole with Schwarzchild radius $r_s$. Its energy in general relativity is
\[
E_{gr} = \sqrt{g_{00}(dt/d\tau)^2} = m\sqrt{\frac{(1-\frac{r_s}{4r})^2}{(1+\frac{r_s}{4r})^2}}.
\]

The relativistic kinetic energy of a test body is
\[
E_{rel} = \frac{m}{\sqrt{1-v^2}}.
\]

The escape velocity is given by setting these equal and solving for $v$ which we call $v_{gr}$,
\begin{equation}
v_{gr} = \sqrt{1 - \frac{(1-\frac{r_s}{4r})^2}{(1+\frac{r_s}{4r})^2}}.
\label{eqn:escape1}
\end{equation}

The event horizon occurs when $v=1$ at $r=r_s/4$. (In Schwarzchild coordinates the event horizon is at $r_s=r$ because of the different radial coordinates.)

In the De Sitter theory a test body above the black hole has the energy,
\[
E_{ds} = \sqrt{G_{00}(dt/d\tau)^2} = m\sqrt{\exp -r_s/r}.
\]

\begin{figure}
\centering
\includegraphics[width = 0.45\textwidth]{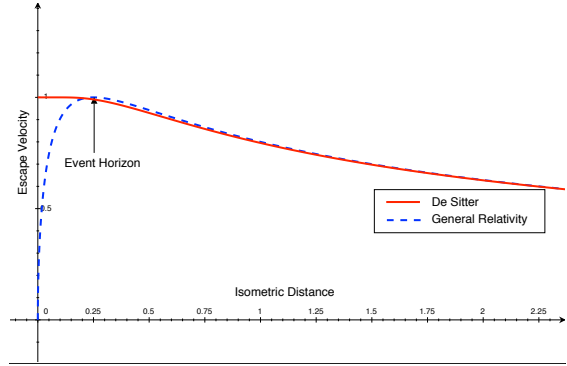}
\caption{The escape velocities in the two theories (\ref{eqn:escape1}) and (\ref{eqn:escape2}) are very close until the test particle approaches the event horizon.}
\label{fig:fig1}
\end{figure}

\begin{figure}
\centering
\includegraphics[width = 0.45\textwidth]{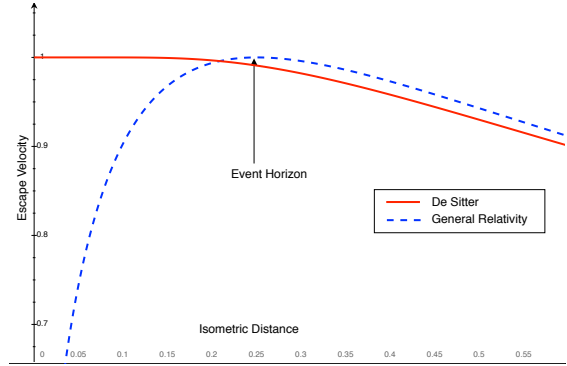}
\caption{Zooming in on the event horizon crossing of Fig. \ref{fig:fig1}: the escape velocity reaches the speed of light in general relativity at the event horizon some distance from the singularity, while the escape velocity in the De Sitter theory remains slightly sublight until the singularity. (The drop in escape velocity past the event horizon in the general relativity solution is a quirk of the isometric coordinate system and not a real phenomenon.)}
\label{fig:fig2}
\end{figure}

The relativistic kinetic energy is the same. Solving for the escape velocity, $v_{ds}$, we have
\begin{equation}
v_{ds} = \sqrt{1 - \exp -r_s/r}.
\label{eqn:escape2}
\end{equation} In this case, $v_{ds}\rightarrow 1$ as $r\rightarrow 0$ monotonically; hence, escape is always possible until the singularity is reached. See Figures \ref{fig:fig1} and \ref{fig:fig2} for an illustration.


\section{Quantum Black Holes: Future Work}
Yang-Mills theory is fundamentally a quantum theory, and of considerable interest is the predictions of De Sitter Yang-Mills theory with respect to quantum black holes. The quantum action allows for Hawking radiation via the conversion of mass energy from the black hole to the surrounding vacuum, allowing virtual vacuum particles to achieve mass shell, become real, and speed away from the black hole as Hawking radiation. This loss of mass energy at a distance results in the black hole evaporating in a well-known process. Hawking radiation is a consequence of thermodynamics that any object with entropy must have thermal emissions. The Yang-Mills theory diverges from the AdS-CFT correspondence theory, however, in that Yang-Mills theory is unitary (given appropriate Faddeev-Popov-DeWitt or BRST formalism to loop sums) without requiring string theory, anti-de Sitter spaces, or other hypotheses outside the Standard Model formalism \cite{Zee:2003}. This means that quantum wavefunctions beginning at infinity (asymptotically flat space), $\phi_i$, and ending at infinity some time $T$ later, $\phi_f$, having their information conserved by the unitary S-matrix of the Yang-Mills theory, $\phi_f = S\phi_i S^{-1}$. In other words, the quantum scattering process of sending matter fields into a black hole or set of black hole-like topologies, and their later emergence loses no information. This issue is not entirely moot in the De Sitter theory, even without the existence of event horizons, because nothing can classically escape the singularities themselves.

\section{Conclusion}
This paper has shown that the De Sitter Yang-Mills theory generates unobserved classical predictions that can be tested in the near future. In particular, while it agrees with GR up to post-Newtonian order, its prediction that black holes have zero-radius event horizons is incompatible with general relativity. An observation that contradicts general relativity such as light emitting from bodies within the predicted event horizon of a black hole would be strong evidence in favor the Yang-Mills theory.

\bibliography{sg2}

\end{document}